\documentclass[twocolumn,prb,amsmath,showpacs,amssymb,amsfonts,10pt]{revtex4-1}

\usepackage{graphicx}
\usepackage{subfigure}
\usepackage[english]{babel}
\usepackage{hyperref}

\begin{document}
\def\bfS{\mathbf{S}}
\def\bfk{\mathbf{k}}
\def\bft{\mathbf{t}}
\def\bfe{\mathbf{e}}
\def\bfx{\mathbf{x}}
\def\tr{\mathrm{tr}}

\title{
First order transition induced by topological defects in the O(3) principal chiral model
}
\author{A.O.~Sorokin}
\email{aosorokin@gmail.com}
\affiliation{Petersburg Nuclear Physics Institute, NRC Kurchatov Institute,\\
188300 Orlova Roscha, Gatchina, Russia}

\date{\today}

\begin{abstract}
Using Monte Carlo simulations, we study thermal and critical properties of two systems, in which domain walls and so-called $Z_2$-vortices as topological defects are presented. The main model is a lattice version of the $O(3)$ principal chiral model. We find a first order transition and give qualitative arguments that the first order is induced by topological defects. We also consider the model of frustrated antiferromagnet on a square lattice with the additional exchange interaction between spins of the third range order. This model belongs to the same symmetry class. In this model, a transition is of first order too.
\end{abstract}


\maketitle

\section{Introduction}

Topological defects play a crucial role in the critical behavior along with conventional perturbative fluctuations. Moreover, one knows examples where a phase transition is driven by topological defects directly. So, in type II superconductors, a transition in the magnetic field is driven by vortex tubes \cite{Abrikosov57}. Another example called now as a topological phase transition is the Berezinskii-Kosterlitz-Thouless (BKT) transition, which occurs in the two-dimensional $O(2)$ model describing XY ferromagnets, and which is driven by vortices \cite{Berezinskii70,Berezinskii71,Kosterlitz73,Kosterlitz74}. A less traditional example is a transition in the Ising model. This model can be entirely reformulated in terms of domain walls and their interaction. Domain walls are line-like topological defects in a two-dimensional model with a discrete order parameter space. So, a phase transition in such models may be considered as a topological one.

In two dimensions, ordinary vortices appear if an order parameter space has the form $G/H=SO(2)\otimes G_{\mathrm{sc}}\otimes G_{\mathrm{d}}$ with $G_{\mathrm{sc}}$ is a simple connected subgroup of a group $G$ and $G_{\mathrm{d}}$ is a discrete subgroup. In a more general case, the criterion of vortices existence is the non-triviality of the first homotopy group $\pi_1(G/H)\neq0$. We know a few classes of physical systems with such properties of an order parameter space, excluding the ordinary case $\pi_1(SO(2))=\mathbb{Z}$. One of such classes is nematics (with a nematic-isotropic transition) described by the classical Maier-Saupe model \cite{Saupe58, Saupe59}. This model is based on the biquadratic interaction $-J(\bfS_{\bfx_1},\bfS_{\bfx_2})^2$, where $\bfS$ is a classical $N$-component vector, so the order parameter space is a real projective space $\mathbb{R}P^{N-1}$ with $\pi_1(\mathbb{R}P^{N-1})=\mathbb{Z}_2$ when $N\geq3$.

Another class of systems with $\pi_1(G/H)=\mathbb{Z}_2$ is the class of frustrated spin systems with isotropic spins ($N=3$) and a non-collinear spin ordering. Such an ordering corresponds to the fully broken symmetry of spin rotations $SO(3)$. As a manifold, the group $SO(3)$ is similar to a 3-sphere $S^3$ with diametrical points being equivalent $SO(3)\approx \mathbb{R}P^3\approx \frac{S^3}{\mathbb{Z}_2}$. Thus, in the both classes, so-called $\mathbb{Z}_2$-vortices appear.

The investigation of thermal and critical properties of two-dimensional systems with $\mathbb{Z}_2$-vortices has a quite long story, since the early '80s. In the works \cite{Solomon81,Duane81,Domany81,Zumbach91,Zumbach92,Pelisetto93}, the possibility of a phase transition at finite temperatures in the $\mathbb{R}P^{N-1}$ model has been discussed. At that, the analysis in work \cite{Duane81} based on the mean field theory and Monte Carlo simulations excludes a transition of a finite order, but not excludes a BKT transition. Similarly, the $SO(3)$ case realized in the model of antiferromagnet on a triangular lattice has been considered in works \cite{Kawamura84,Kawamura93,Southern93,Wintel94,Southern95,Wintel95}, where a BKT-type transition has been predicted. Most of the works (in the both cases) use Monte Carlo simulations and show the presence of a singularity in thermal behavior typical to a phase transition.

However, there are arguments against the existence of a phase transition at a finite temperature based on the $\sigma$-model approach \cite{Hasenbusch96,Niedermayer96,Azaria01}. The $\sigma$-model is the effective theory describing low-energy (weak) fluctuations, so it also describes a low-temperature behavior. Due to the weakness of fluctuations, an interaction between them depends only on a local geometry of the order parameter space and does not feel a topology of the space $G/H$ \cite{Friedan80,Friedan85}. Thus, the cases of $G/H=SO(3)$ and $\mathbb{R}P^3$ are equivalent to the case $G/H=S^3=O(4)/O(3)$ and have the same low-temperature behavior as the $O(4)$ model \cite{Hikami81,Azaria90,Azaria93}. In two dimensions, the $\sigma$-model with a non-flat space $G/H$ predicts the absence of long-range (the Mermin-Wagner theorem) or quasi-long-range orders at a finite temperature and exponential decrease of the correlation length with temperature increasing. So, a transition-like behavior observed numerically can be explained only as a crossover between the $\sigma$-model behavior and the high-temperature behavior with an appreciable density of $\mathbb{Z}_2$-vortices \cite{Azaria01}. In the recent work \cite{Sinner14}, the crossover in the $SO(3)$ principal chiral model have been observed by the non-perturbative renormalization group (RG) approach. Also, the $\mathbb{Z}_2$-vortex concept is used to explain an anomalous behavior of some triangular antiferromagnets, observed experimentally (see \cite{Kawamura10,Kawamura10-2,Kawamura11} and refs. therein).

In this paper we consider the possibility of participation of $\mathbb{Z}_2$-vortices in a {\it bona fide} topological phase transition. Of course, for this we need to take a model with a more complicated order parameter space. We investigate two models with $G/H=O(3)\equiv\mathbb{Z}_2\otimes SO(3)$. There are two types of topological defects presented in this symmetry class, $\mathbb{Z}_2$-vortices and domain walls. We expect that an interaction of these defects allows $\mathbb{Z}_2$-vortices influence a critical behavior.

We have already known the case when an interaction between two types of topological defects changes a critical behavior, and this case has served us as a hint. This is the case of the Ising-$O(2)$ model with $G/H=\mathbb{Z}_2\otimes SO(2)$, where (ordinary) vortices and domain walls are presented too (see \cite{Korshunov06} for a review).
Accurate analysis of numerical results for different systems of this class allows to formulate two possible scenarios: either a BKT transition occurs at temperature below an Ising transition, or these transitions occurs at the same temperature as a first order transition \cite{Olsson95,Olsson97,Korshunov02}.

Korshunov argued \cite{Korshunov02} that the first scenario is possible in systems where fractional vortices are present in the spectrum of topological defects. Fractional vortices appear as some kinds of kinks propagating on domain walls. The logarithmical interaction of these kinks is weaker than the interaction of the conventional vortices and leads to a phase transition on a domain wall at $T_{\rm k}<T_{\rm BKT}$. At $T>T_{\rm k}$, the domain wall turns opaque for the correlations of the $SO(2)$ parameter. As a consequence, on approaching the continuous Ising-like transition, the quasi-long-range $SO(2)$ order has to break down, and the BKT-transition has to occurs at $T_{\rm BKT} < T_{\rm Is}$. Such fractional vortices are found for many models from the class of the Ising-$O(2)$ model: the fully frustrated XY model \cite{Halsey85,Korshunov86-2}, XY antiferromagnet on a triangular lattice \cite{Korshunov86}, XY helimagnets \cite{Uimin91,Uimin94}, etc.

The second scenario when two transitions coincide is also observed in the Ising-XY model \cite{Granato91,Granato91-2,Granato95,Hasenbusch05,Hasenbusch05-2} and XY J$_1$-J$_3$ model on a square lattice \cite{Sorokin12}. (The $N=3$ case of the last model is considered in the current study, see the description of the model below.) This single transition is of first order.

Of course, the analogy with the case of the Ising-$O(2)$ model can not be complete for two reasons. Firstly, the group $SO(3)$ is non-Abelian, so perturbative excitations can not be integrated out unlike to the Abelian $SO(2)$ case, and we can not formulate the model in terms of topological defects. Secondly, properties of $\mathbb{Z}_2$-vortices are very different from usual vortices, in particular fractional vortices do not exist. Nevertheless, in this paper we demonstrate that $\mathbb{Z}_2$-vortices and domain wall interact and lead to a single first order transition.

In frustrated spin systems, the coset $G/H=\mathbb{Z}_2\otimes SO(3)$ as an order parameter space appears in several ways. Generally speaking, a spin lattice model has the symmetry $O(N)\otimes G_l$, where $G_l$ is a discrete lattice symmetry. If $N=3$, one has two possibilities. Firstly, a spin ordering is non-planar, so the full symmetry of spin rotations and inversion $O(3)$ is broken. And secondly, a spin ordering is planar, and a inversion symmetry (of a spin space) remains unbroken, but $\mathbb{Z}_2$ subgroup of a lattice group is also broken. The second case is often accompanied by the "order from disorder"\ phenomenon.

A few models of frustrated spin system with the $\mathbb{Z}_2\otimes SO(3)$ order parameter space have been considered in works \cite{Sachdev04,Domenge08,Tamura08,Kawashima10,Tamura11,Tamura13}. In the work \cite{Sachdev04}, the J$_1$-J$_3$ model on a square lattice have been considered. (The expression J$_1$-J$_3$ means that we deal with a model of antiferromagnet with competing interaction between nearest spins and spins of the third range order.) The authors have found a second order phase transition with exponents of the Ising model. In this work, we also consider this model and find a first order transition, that is discussed below. A second order transition have been also found in a special case of the J$_1$-J$_3$ model on a triangular lattice \cite{Tamura13}. But the rest works have shown the first order of a transition in the J$_1$-J$_2$ model on a kagome lattice \cite{Domenge08}, J$_1$-J$_2$ model on a honeycomb lattice \cite{Kawashima10}, and J$_1$-J$_3$ model on a triangular lattice \cite{Tamura08,Tamura11}.

Beside the J$_1$-J$_3$ model on a square lattice, we consider numerically two matrix models on a square lattice which directly realize the $\mathbb{Z}_2\otimes SO(3)$ and $SO(3)$ order parameter spaces. To reveal an interaction between vortices and domain walls, the $SO(3)$ case is also considered and compared with the $\mathbb{Z}_2\otimes SO(3)$ case.

\section{Models and method}

A non-planar spin ordering is described by three orthogonal $N$-component vectors. Generally, a set of orientations of $P$ orthogonal vector in $N$ dimensions is the Stiefel manifold \cite{Stiefel35}
\begin{equation}
    V_{N,P}=\frac{O(N)}{O(N-P)},
    \label{manifold}
\end{equation}
with the special cases
\begin{equation}
    V_{1,1}=\mathbb{Z}_2,\quad V_{N,1}=S^{N-1},
\end{equation}
\begin{equation}
    V_{N,N-1}=SO(N), \quad V_{N,N}=\mathbb{Z}_2\otimes SO(N).
\end{equation}
We are interested in the cases $N=3$ and $P=2,\,3$. The order parameter is a $3\times P$ matrix composed of $P$ orthogonal $3$-vectors
\begin{equation}
    \Phi\left(V_{3,2}\right)=(\bfS,\bfk),\quad \Phi\left(V_{3,3}\right)=(\bfS,\bfk,\bft).
\end{equation}
The discrete form of the $\sigma$-model is \cite{Zumbach93}
\begin{equation}
    H=-J\sum_{\bfx,\mu}\tr\,\Phi_\bfx^T\Phi_{\bfx+\bfe_\mu},\quad \mu=1,\,2,
\end{equation}
where $\bfe_\mu$ is a unit vector of a square lattice, $J>0$. In a general case, the Hamiltonian is invariant under the group $O(N)\otimes O(P)$, where $L$ and $R$ mean the left and right action of a rotation matrix on the order parameter $\Phi$. When a ground state configuration is chosen, the symmetry is broken to the $O(N-P)_L\otimes O(P)_D$ subgroup with $O(D)$ acting simultaneously both right and left (diagonal subgroup). Thus, one see that the order parameter space is \eqref{manifold}. In particular, the $V_{3,3}$ Stiefel model is equivalent to the $O(3)_L\otimes O(3)_R$ (principal) chiral model.

In simulations we use following definitions of the order parameter
\begin{equation}
    m=\sum_\bfx \bfS_\bfx, \quad \bar m=\sqrt{\langle m^2\rangle},
\end{equation}
\begin{equation}
    k=\sum_\bfx \bfk_\bfx, \quad \bar k=\sqrt{\langle k^2\rangle}
\end{equation}
for the $SO(3)$ sector. And
\begin{equation}
    \sigma=\sum_\bfx\sigma_\bfx=\sum_\bfx \det \Phi_\bfx, \quad \bar \sigma=\sqrt{\langle |\sigma|\rangle}
\end{equation}
for the $\mathbb{Z}_2$ sector of the $V_{3,3}$ model. We monitor the first, second and fourth moments of the order parameters $p=m,\,k,\,\sigma$ and internal energy density $E$, to have information on the specific heat $C$, susceptibilities $\chi_p$ and higher order cumulants, e.g. the Binder's cumulant.

We also compute the helicity modulus, because at low temperatures its size dependence is the most convincing evidence for the validity of the $\sigma$-model prediction.
$$
\Upsilon_{\mu,a}=\frac{1}{L^2}\left\langle\sum_\bfx\left[\bfS_\bfx^b\cdot\bfS_{\bfx+\bfe_\mu}^b+\bfS_\bfx^c\cdot\bfS_{\bfx+\bfe_\mu}^c\right]\right\rangle-
$$
\begin{equation}
\frac{1}{TL^2}\left\langle\left(\sum_\bfx\left[\bfS_\bfx^b\cdot\bfS_{\bfx+\bfe_\mu}^c-\bfS_\bfx^c\cdot\bfS_{\bfx+\bfe_\mu}^b\right]\right)^2\right\rangle,
\label{helicity}
\end{equation}
\begin{equation}
    \Upsilon_\mu=\frac13\sum_a\Upsilon_{\mu,a},
\end{equation}
where $L$ is a lattice size. Note that at zero temperature $\Upsilon_{\mu,1}=0$ and $\Upsilon_{\mu,2}=\Upsilon_{\mu,3}=1$.
\begin{figure}[t]
\includegraphics[scale=0.45]{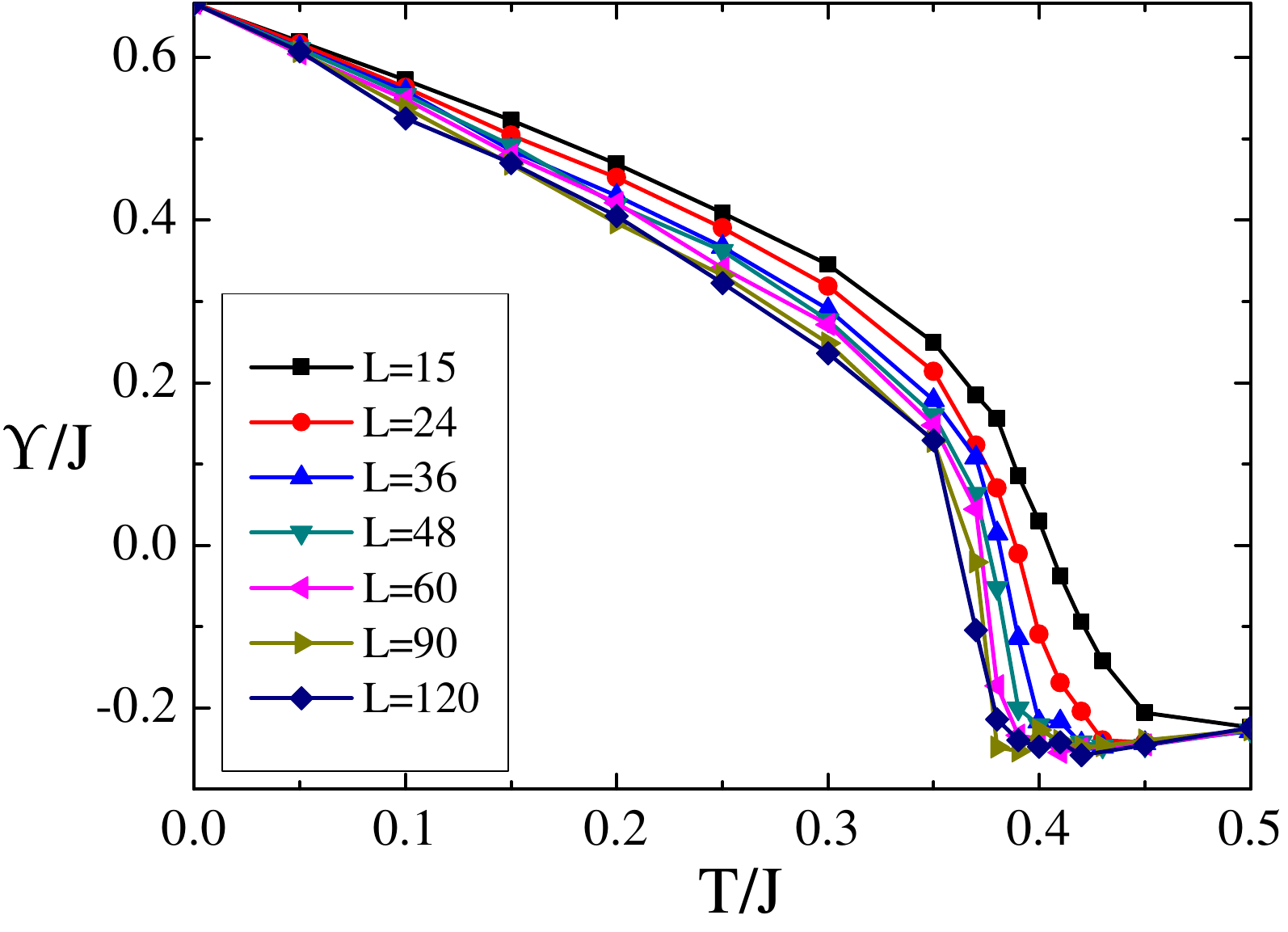}
\includegraphics[scale=0.45]{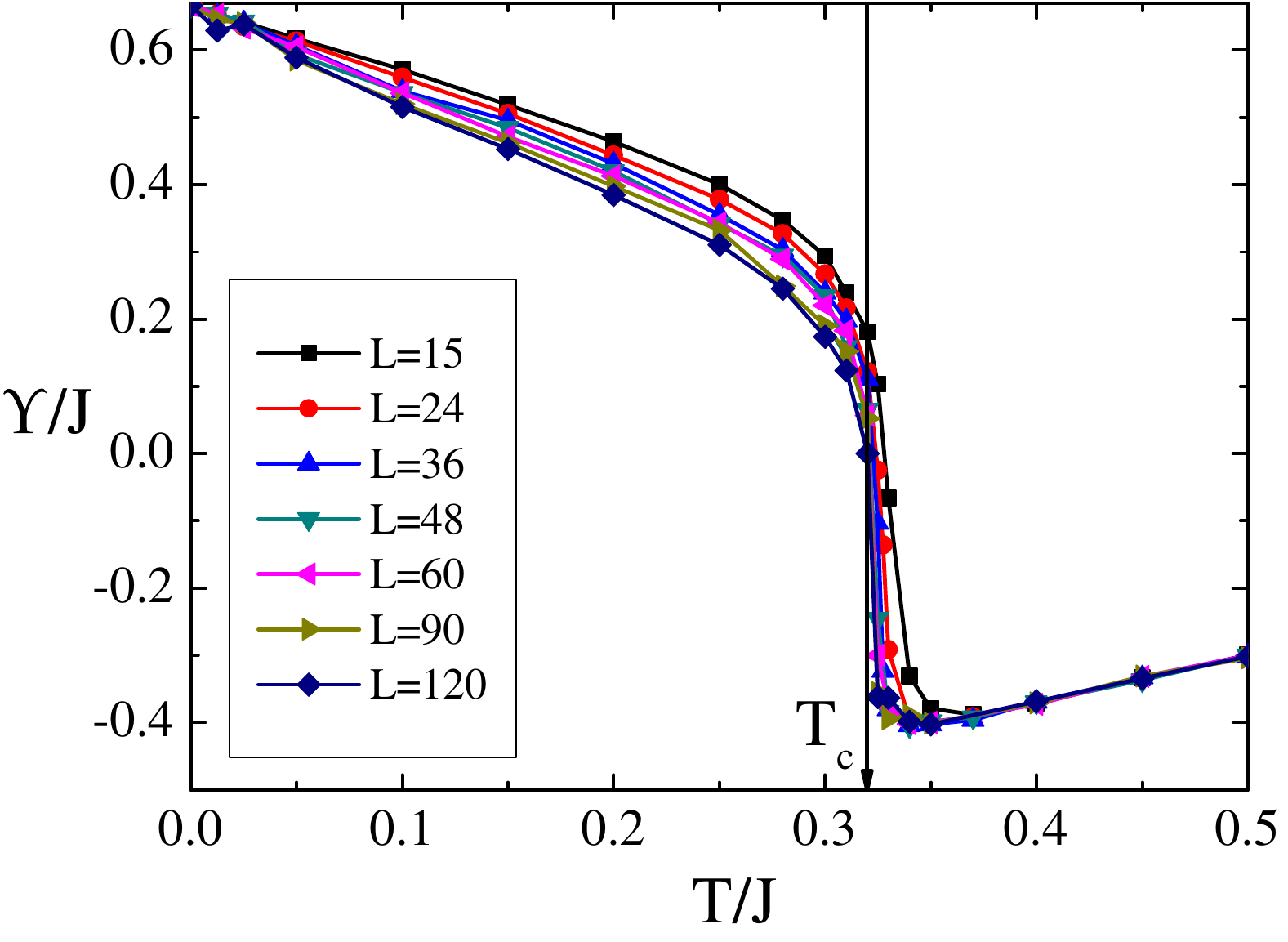}
\caption{\label{V32-Y}Thermal dependence of the helicity modulus in the $V_{3,2}$ and $V_{3,3}$ Stiefel models}
\end{figure}

The definition of $\mathbb{Z}_2$-vortices is following. It is known that the group $SO(3)$ is not simply connected, and its (double) covering group is $SU(2)$. So, an element $\Phi$ of $SO(3)$ corresponds to two elements of $SU(2)$, namely $U$ and $-U$. Consider a changing of the order parameter along a closed contour $\mathcal C$
\begin{equation}
    \Psi_{\mathbf{xx}'}=\Phi_\mathbf{x}^{-1}\Phi_{\mathbf{x}'}.
\end{equation}
\begin{equation}
    \Psi_{\mathcal{C}}=\left(\prod_{i=1}^n\Psi_{\mathbf{x}_i\mathbf{x}_{i+1}}\right)\Psi_{\mathbf{x}_n\mathbf{x}_1}=1.
\end{equation}
Using the homomorphism $f:SO(3)\to SU(2)$, we define $V_{\mathbf{xx}'}\equiv f(\Psi_{\mathbf{xx}'})=f(\Phi_\mathbf{x}^{-1}\Phi_{\mathbf{x}'})=    f(\Phi_\mathbf{x}^{-1})f(\Phi_{\mathbf{x}'})=U_\mathbf{x}^{-1}U_{\mathbf{x}'}$. In particular, $V_{\mathcal{C}}=\pm1$. An unitary matrix $V(\Psi)$ can be constructed using the parametrization of an orthogonal matrix $\Psi$ by Euler angles $\varphi,\,\theta,\,\psi$, and then
\begin{equation}
    V=e^{\frac i2\varphi\sigma_3} e^{\frac i2\theta\sigma_1} e^{\frac i2\psi\sigma_3}
\end{equation}
Therefore the vorticity inside a primitive cell is
\begin{equation}
 v_\bfx=\frac12\left(1-\frac12\mathrm{Tr}\prod_\square V\right).
\end{equation}
And the density (concentration) of vortices is
\begin{equation}
    \rho_v=\frac{1}{L^2}\sum_\mathbf{x}v_\mathbf{x}.
\end{equation}
Note that the order parameter $\Phi$ of the $V_{3,2}$ model can be easily extended to a $3\times3$ matrix by adding the vector $\bft=\bfS\times\bfk$.
The density of domain walls is defined simpler
\begin{equation}
    w_{\bfx,\mu}=\frac12\left(1-\sigma_\bfx\sigma_{\bfx+\bfe_\mu}\right),\quad
    \rho_w=\frac{1}{2L^2}\sum_{\bfx,\mu}w_{\bfx,\mu}.
\end{equation}
For the density of topological defects, we also calculate the analogue of a susceptibility, called as the topological susceptibility
\begin{equation}
    \chi_{\mathrm{td}}=L^2\left(\langle\rho_{\mathrm{td}}^2\rangle-\langle\rho_{\mathrm{td}}\rangle^2\right),
\end{equation}
where the subscript "td"\ means topological defects --- vortices and domain walls. It is expected that this quantity has a singularity at a critical point.

To study the models, we use extensive Monte Carlo simulations based on the over-relaxed algorithm \cite{Brown87, Creutz87}. To define the order of a transition, we use the histogram analysis method. Thermalization is performed within $3\cdot10^5$ Monte Carlo steps per spin, and calculation of averages, within $3.3\cdot10^6$ steps. We use periodic boundary conditions and consider lattices with sizes $15\leq L\leq 120$.

\section{$V_{3,2}$ and $V_{3,3}$ Stiefel models}

\subsection{Low-temperature behavior}

\begin{figure}[t]
\includegraphics[scale=0.45]{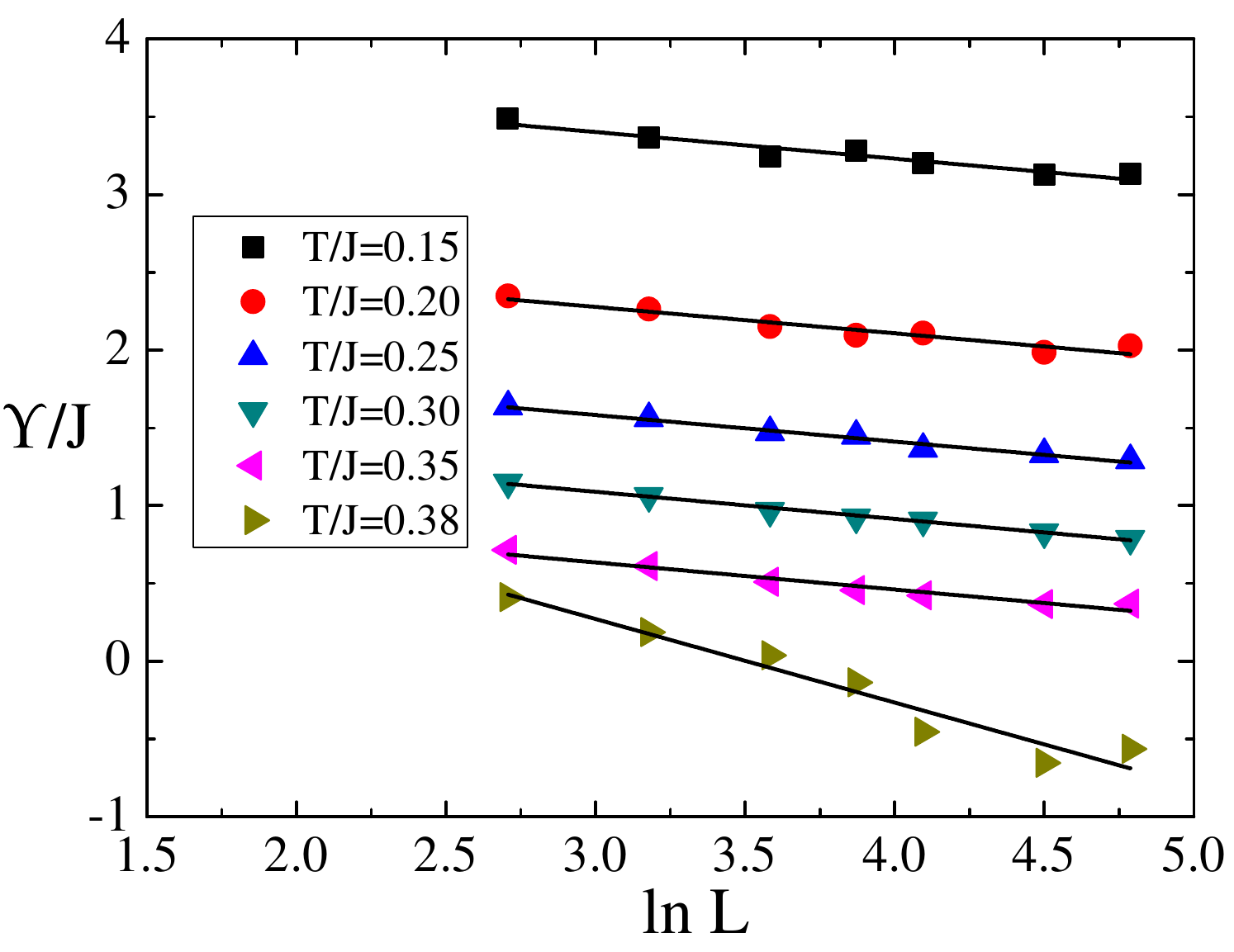}
\includegraphics[scale=0.45]{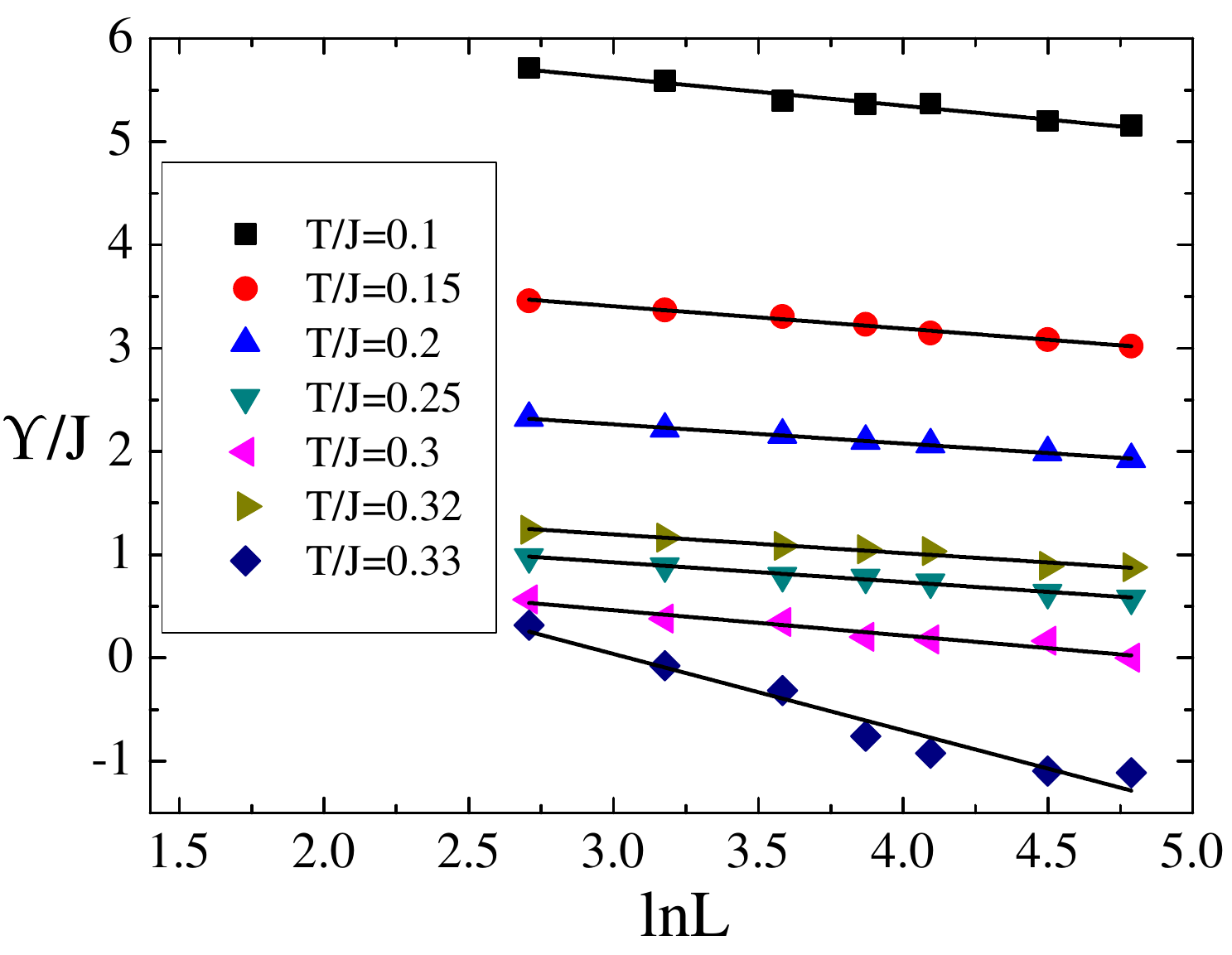}
\caption{\label{V32-YlnL}Lattice size dependence of the helicity modulus in the $V_{3,2}$ and $V_{3,3}$ models.}
\end{figure}

\begin{figure}[t]
\includegraphics[scale=0.45]{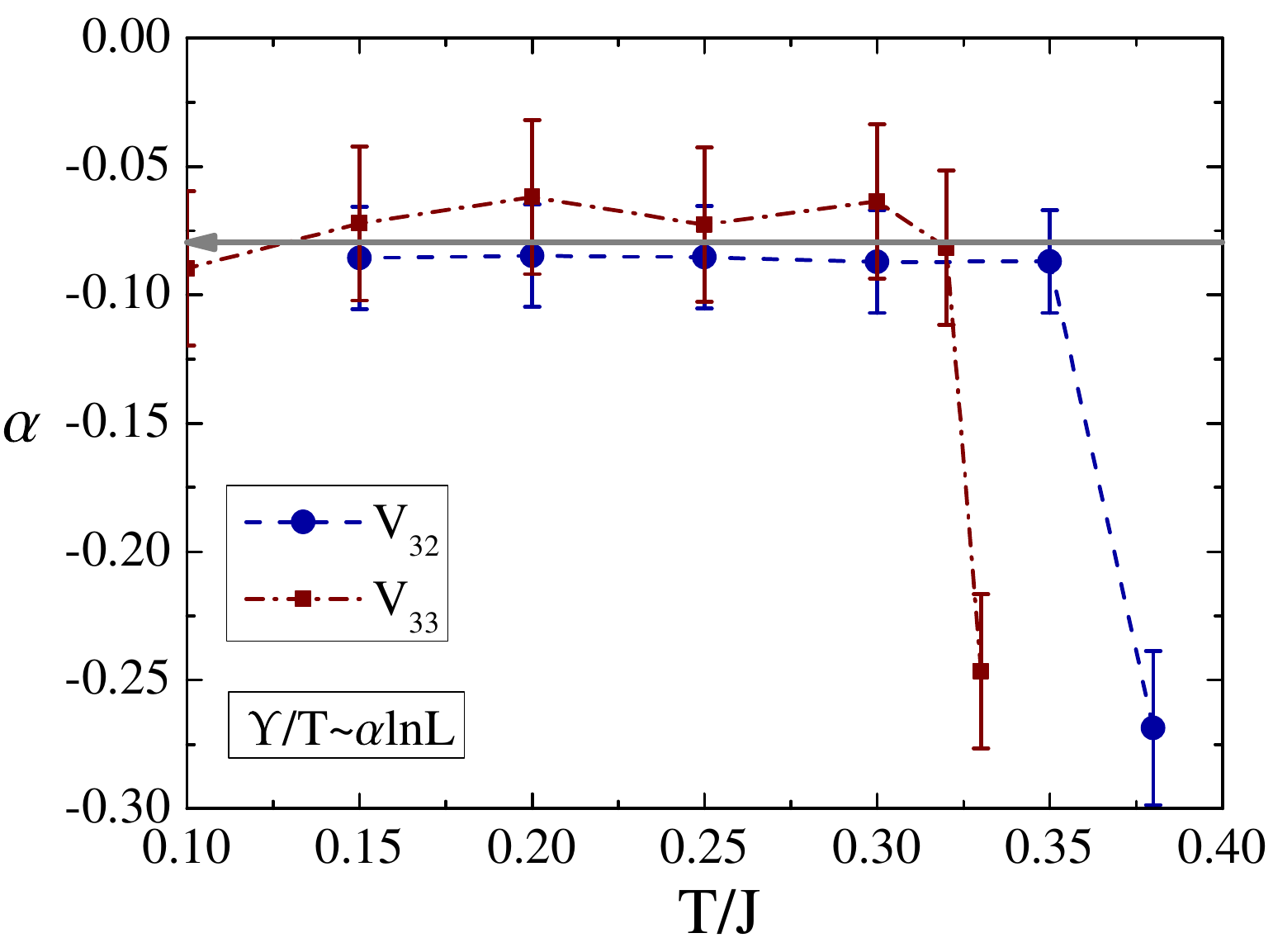}
\caption{\label{V3P-Slope}Comparison of the lattice size dependence of the helicity modulus in the $V_{3,2}$ and $V_{3,3}$ models with the $\sigma$-model result. The grey line marks the value $\frac1{4\pi}$.}
\end{figure}

As we have announced above, the finite-size scaling dependence of the helicity modulus al low temperatures may be compared with the prediction of the $\sigma$ model RG-calculation \cite{Azaria92}
\begin{equation}
    \frac{\Upsilon(L)}{T}\sim\frac1{4P\pi}\ln\left(\frac{\xi}{L}\right),
    \label{azaria}
\end{equation}
where the factor $P$ appears due to we calculate the helicity modulus only for the vector $\bfS$ without the vectors $\bfk$ and $\bft$.
In contrast to the $O(2)$ model, where the helicity modulus remains non-zero at all temperatures below a BKT transition and has imperceptible
finite-size scaling corrections, we find that this quantity tends to zero with lattice size increasing (fig. \ref{V32-Y}). This indicates the absence of a quasi-long-range order in the $SO(3)$ parameter. The helicity modulus dependence on a size $\Upsilon(L)$ is in a good agreement with formula \eqref{azaria} for the $V_{3,2}$ model as well as for the $V_{3,3}$ model (figs. \ref{V32-YlnL} and \ref{V3P-Slope}).
In other words, these models have very similar low-temperature behavior, at that the behavior is the same as in the $O(4)$ model, where no vortices or domain walls.

The $\sigma$-model behavior of the helicity modulus is observed in a wide range of temperature (fig. \ref{V3P-Slope}). But close to the temperature of the transition or crossover, a character of thermal and lattice dependence of $\Upsilon$ drastically changes.

\subsection{Crossover in the $V_{3,2}$ Stiefel model}

\begin{figure}[t]
\includegraphics[scale=0.45]{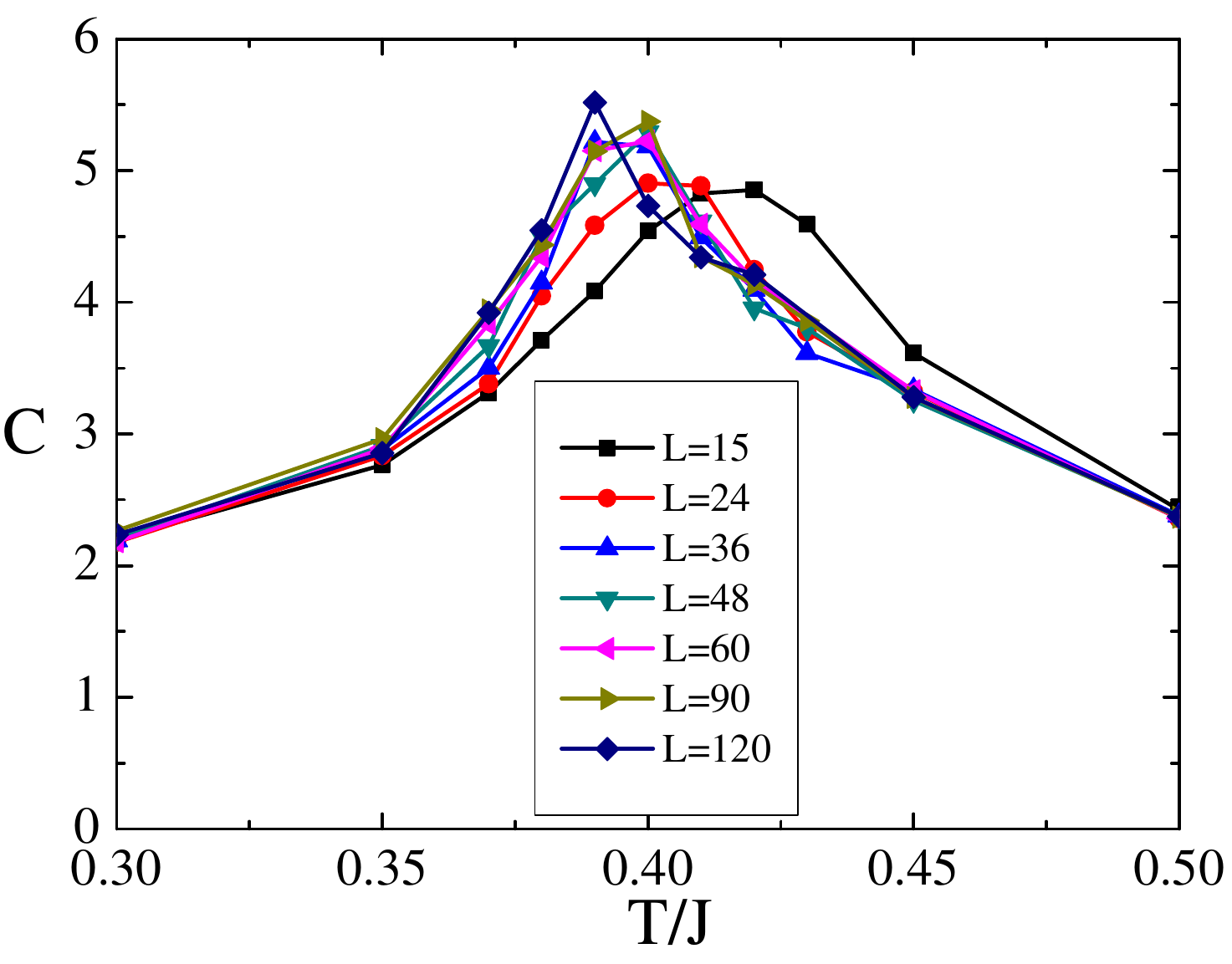}
\includegraphics[scale=0.45]{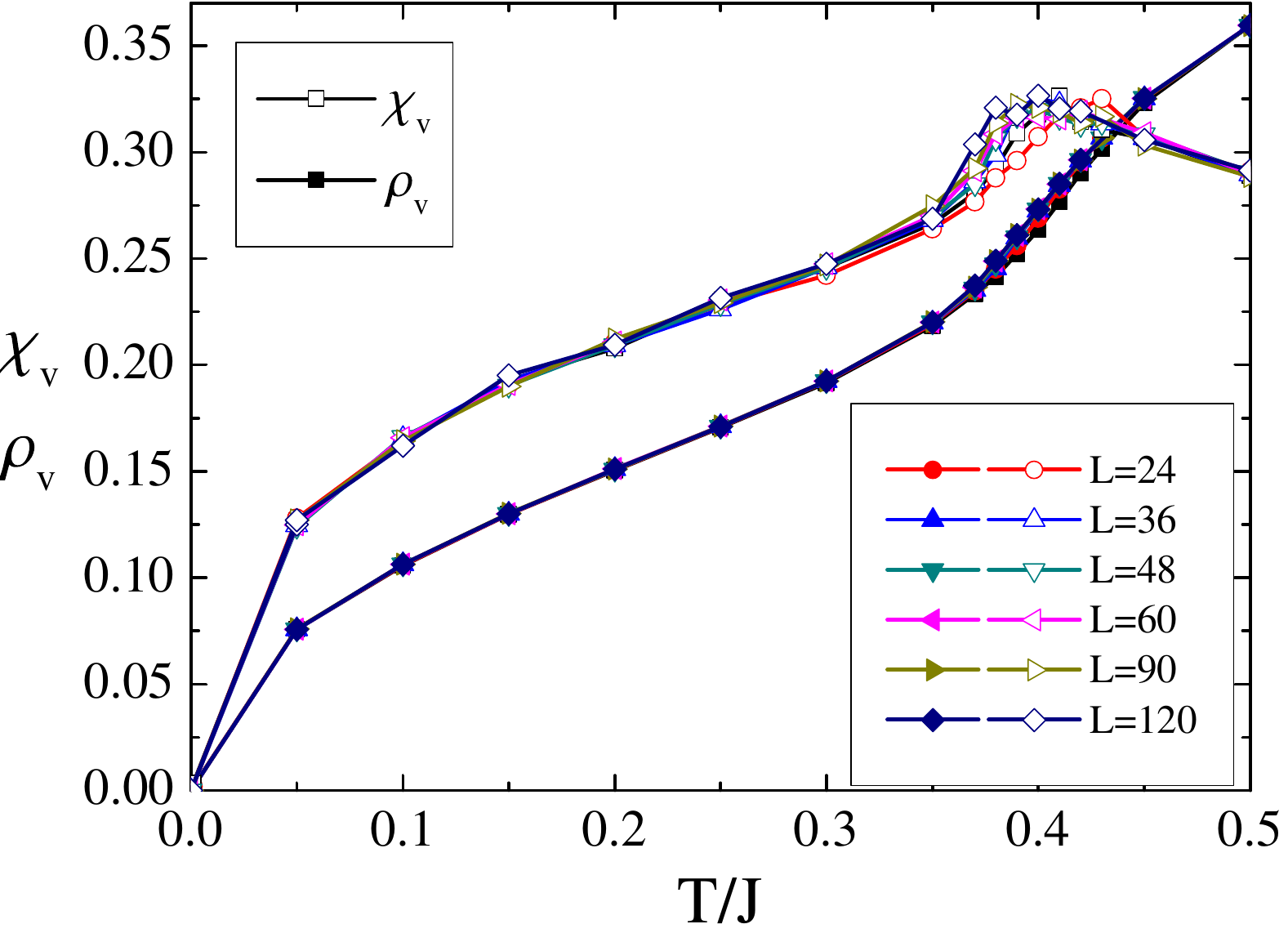}
\caption{\label{V32-vx}Thermal dependence of the specific heat, topological density and susceptibility in the $V_{3,2}$ Stiefel model}
\end{figure}

The crossover induced by $\mathbb{Z}_2$-vortices manifests as features of the thermal dependence of thermodynamic quantities. We have discussed the changes in the behavior of $\Upsilon$, which implies a change in the thermal dependence of the correlation length $\xi$, according to the formula \eqref{azaria}. Similar features are observed for other quantities at the same value of temperature
\begin{equation}
    \frac TJ=0.39(1).
\end{equation}
The specific heat and topological susceptibility have a peak at the crossover temperature (see fig. \ref{V32-vx}). This peak is not evidence of a singularity. We expect that values of these quantities remain finite in the thermodynamical limit $L\to\infty$.

In fact, peaks (or very weak singularities) of these quantities exclude the possibility that this crossover is a BKT transition, since such features is not observed upon a genuine BKT transition. Another difference consists in a fact that below the crossover temperature the density of vortices and susceptibility decrease much slower than that is observed for the $O(2)$ model. In other words, the process of association in the pair for $\mathbb{Z}_2$-vortices is much less noticeable than in the case of ordinary vortices.

As an analogy, this crossover is reminiscent of a crossover in a supercritical fluid in a liquid-gas phase diagram. Within this analogy, the density of vortices serves as an order parameter.
\begin{figure}[t]
\includegraphics[scale=0.45]{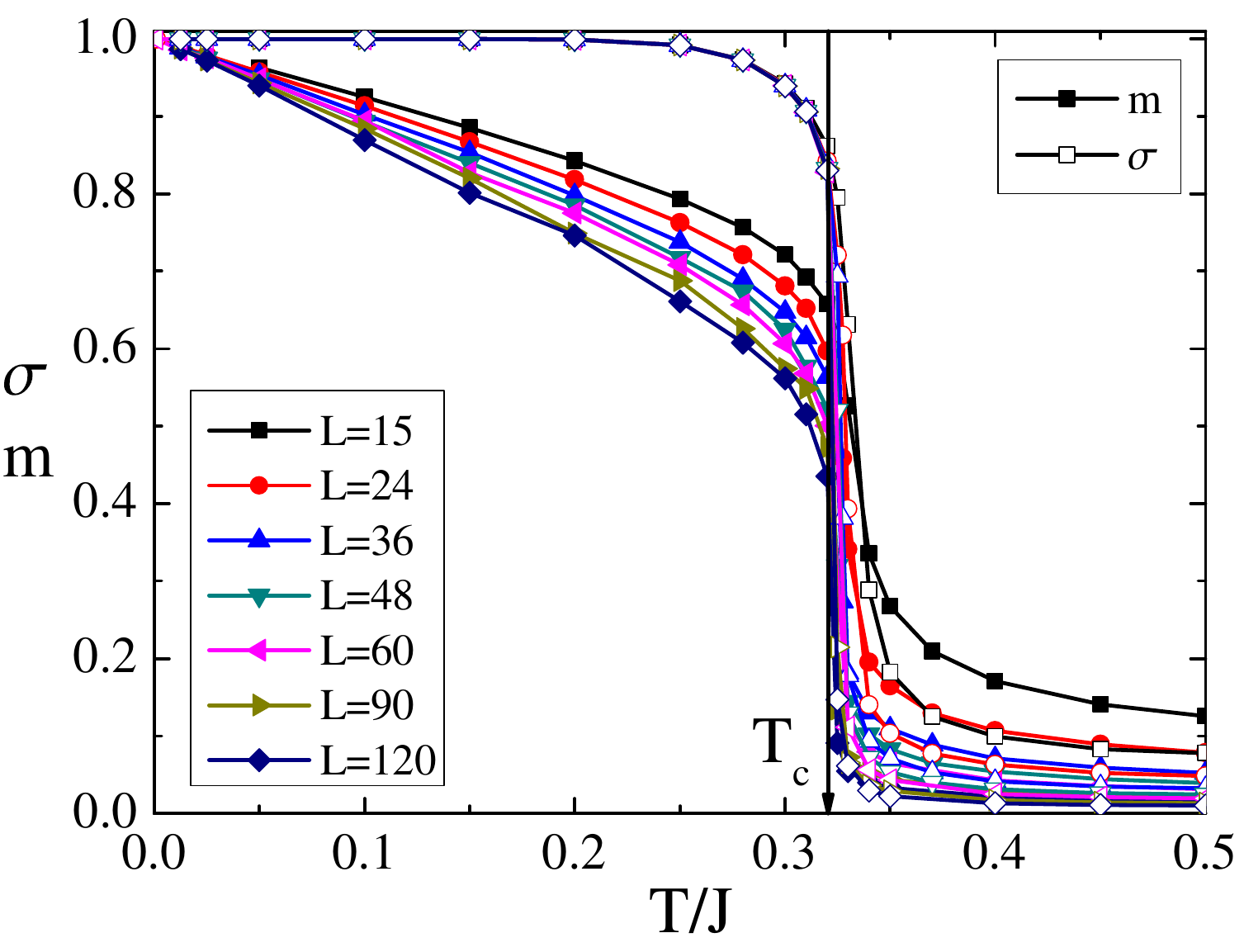}
\includegraphics[scale=0.45]{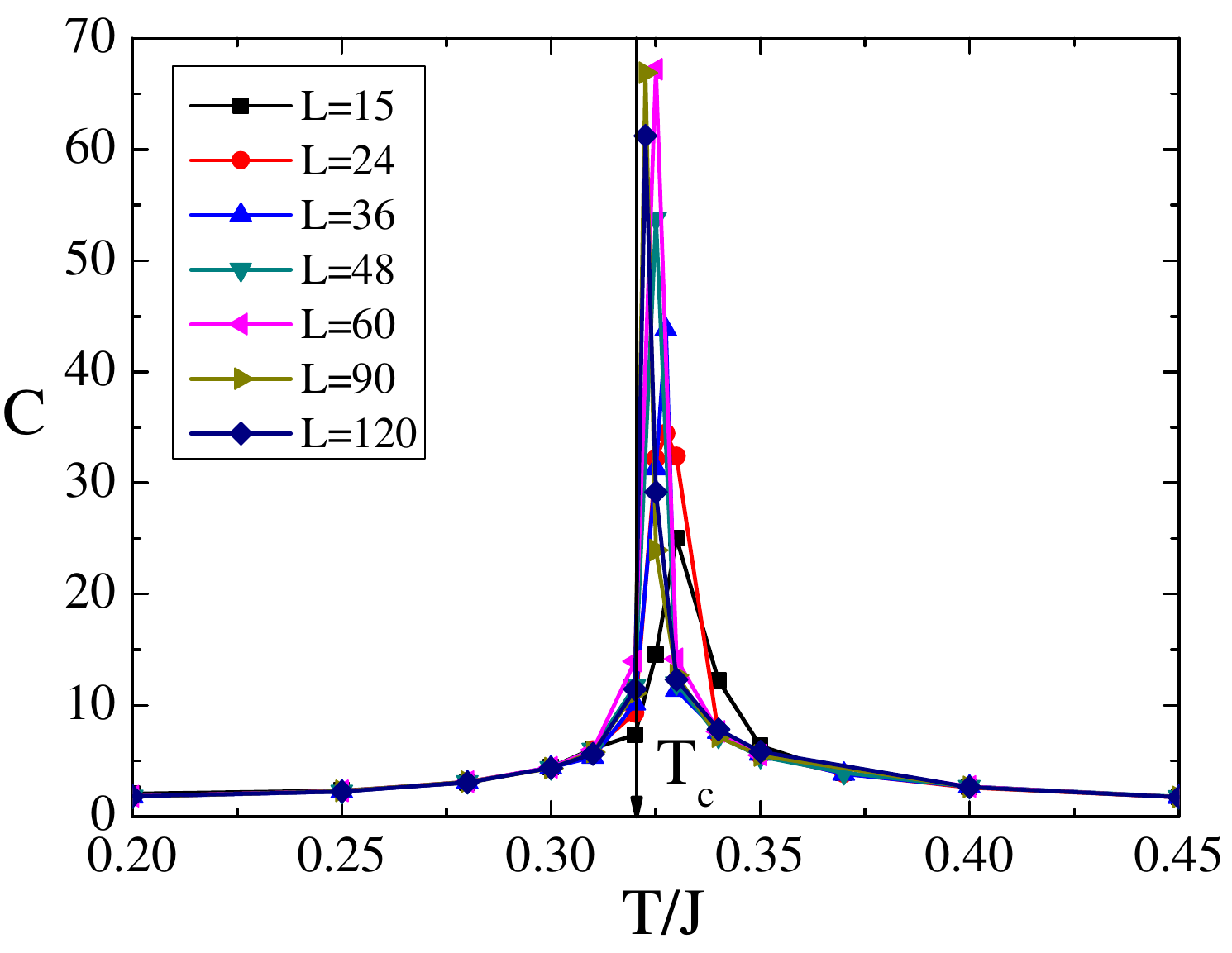}
\caption{\label{V33-m}Thermal dependence of the order parameters and specific heat in the $V_{3,3}$ Stiefel model.}
\end{figure}

\subsection{First order transition in the $V_{3,3}$ Stiefel model}

\begin{figure}[t]
\includegraphics[scale=0.45]{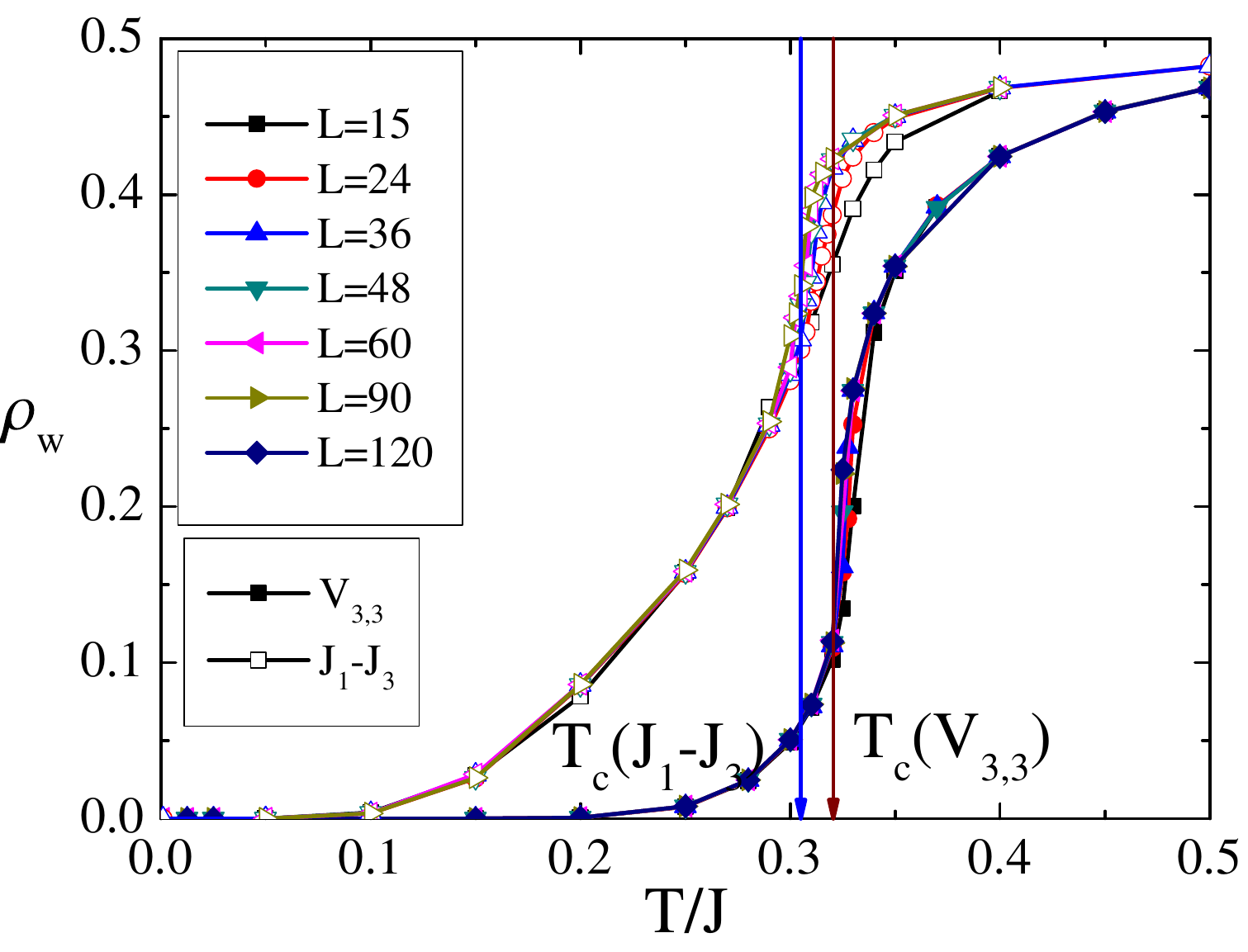}
\includegraphics[scale=0.45]{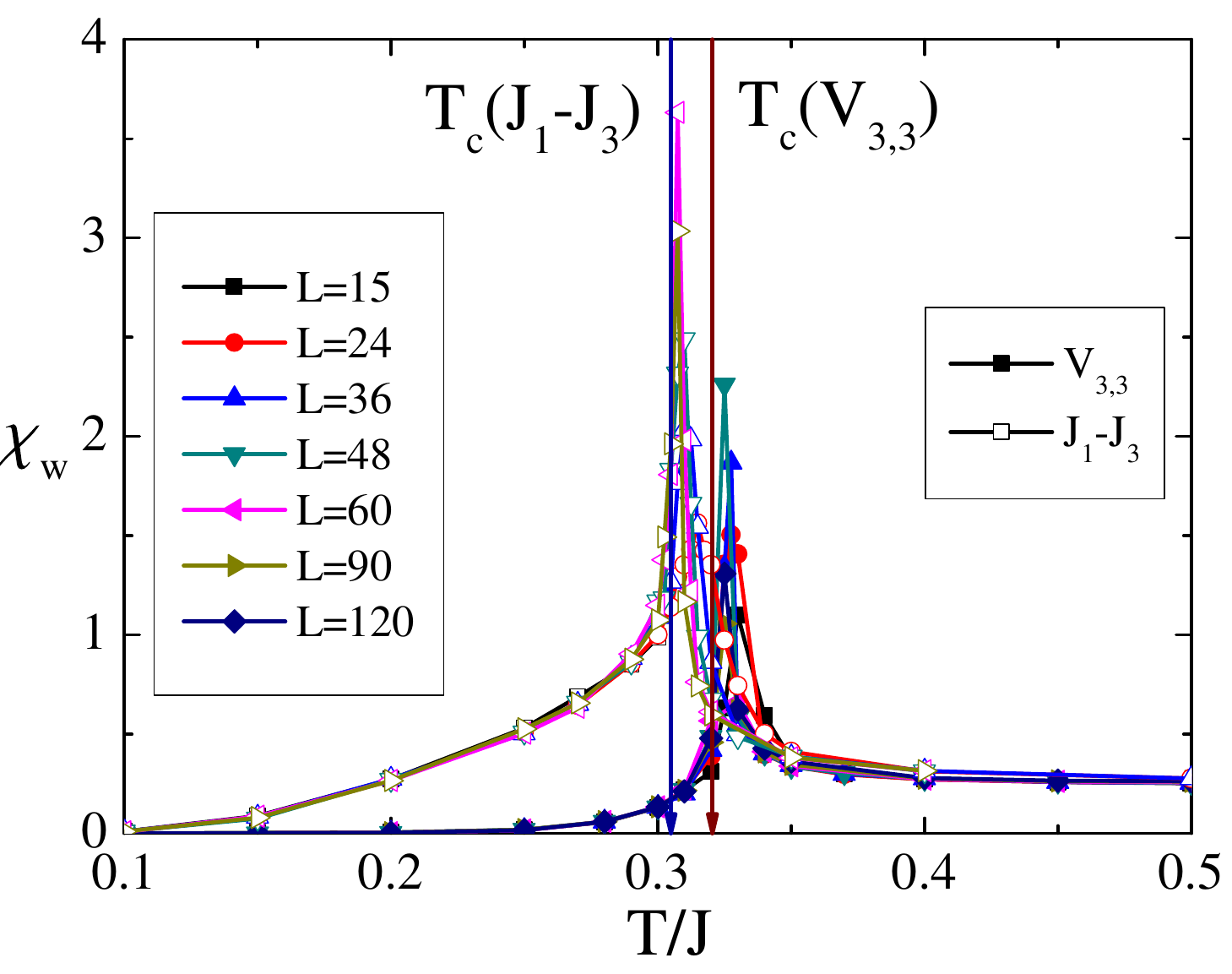}
\caption{\label{V33-wx}Thermal dependence of the density of domain walls and susceptibility in the $V_{3,3}$ Stiefel and J$_1$-J$_3$ model.}
\end{figure}
\begin{figure}[t]
\includegraphics[scale=0.45]{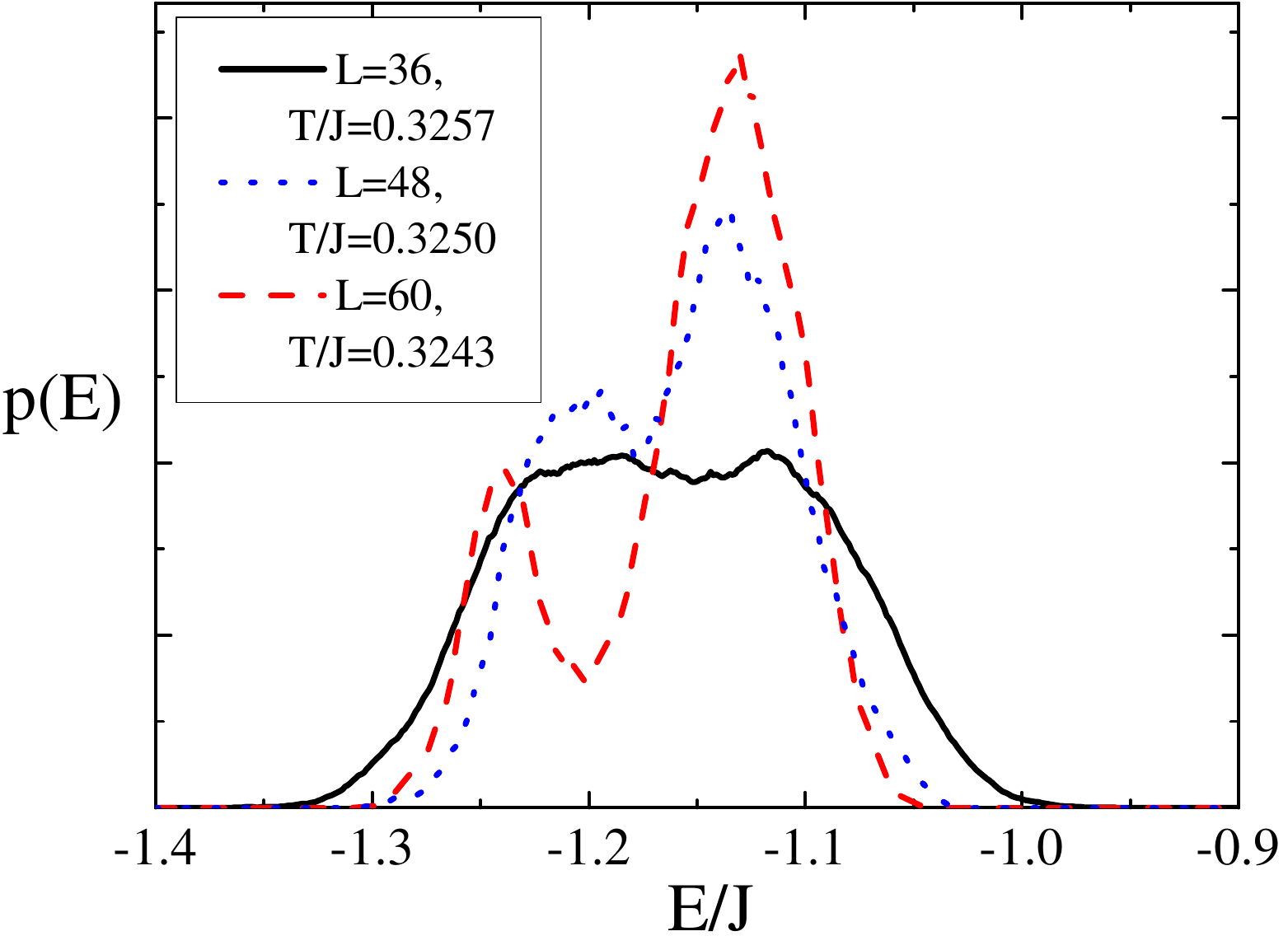}
\includegraphics[scale=0.45]{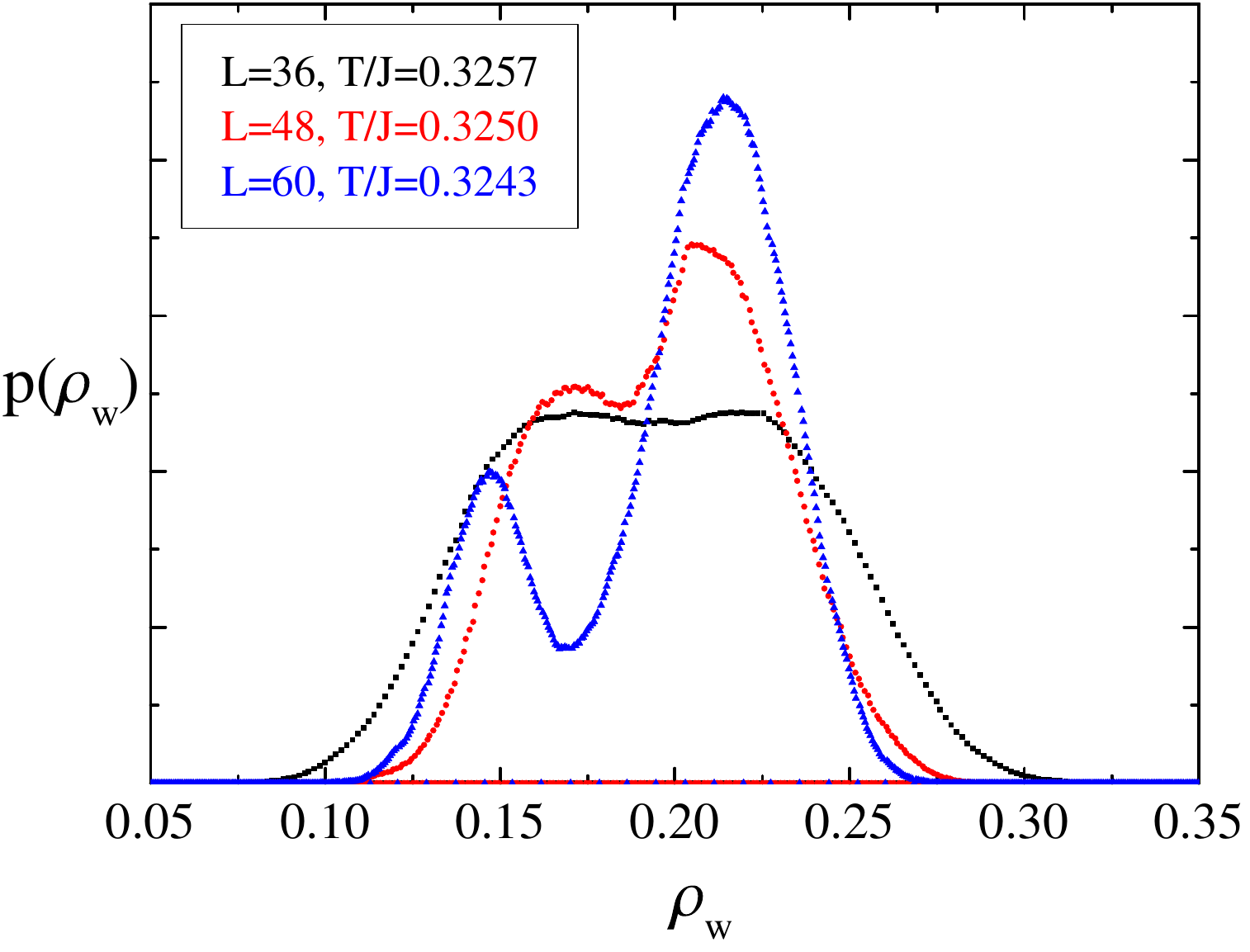}
\caption{\label{V33-pw}Energy and walls density distributions close to the transition point in the $V_{3,3}$ model.}
\end{figure}
\begin{figure}[t]
\includegraphics[scale=0.45]{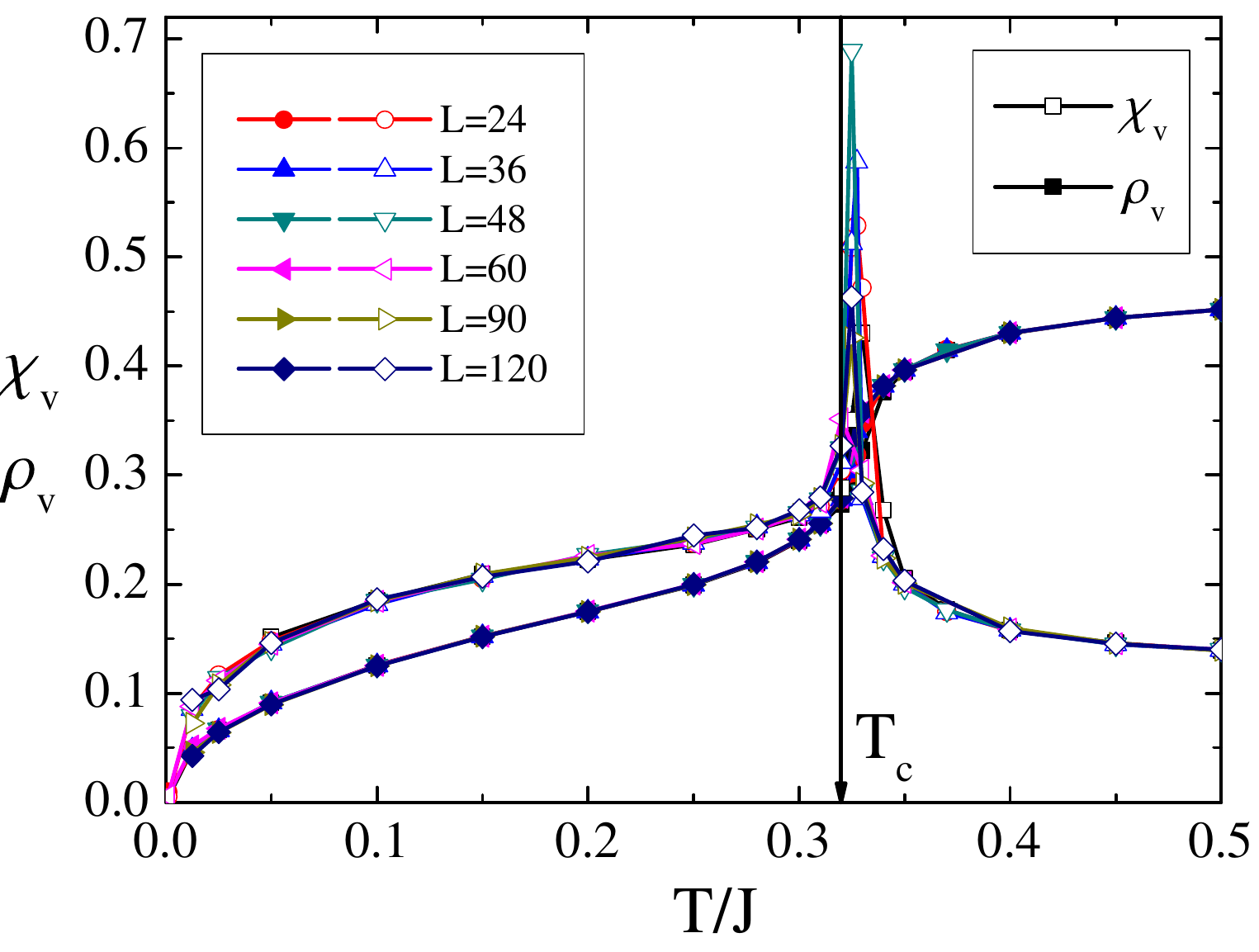}
\caption{\label{V33-vx}Thermal properties of $\mathbb{Z}_2$-vortices in the $V_{3,3}$ model.}
\end{figure}

In contrast to the $V_{3,2}$ model, the $V_{3,3}$ model has the additional discrete symmetry. No-go theorems forbidding a spontaneously breaking of a global discrete symmetry in two dimensions do not exist. And really, we observe the long-range order in the parameter $\sigma$ below the transition temperature, $\bar\sigma\neq0$, while the $SO(3)$ order parameters vanish $\bar m=\bar k=0$ (in the limit $L\to\infty$) at any non-zero temperature (fig. \ref{V33-m}).

At temperature
\begin{equation}
    \frac{T_c}J=0.3201(5),
\end{equation}
the phase transition occurs. Our data favor the first order of the transition. The specific heat (fig. \ref{V33-m}) has a singularity stronger than it is expected upon a second order Ising transition ($\frac{\alpha}{\nu}\approx2$ instead of $\frac{\alpha}{\nu}=0$). This observation applies also to the topological (domain walls) susceptibility (fig. \ref{V33-wx}). We remind that in the pure Ising model on a square lattice, the internal energy density relates to the domain walls density as $E=-2+4\langle\rho_w\rangle$, so the specific heat and topological susceptibility have the same (logarithmical) singularity $C\sim\chi_w\sim\ln(T-T_c)$. But fig. \ref{V33-wx} shows a more singular behavior.

The most important criterion for determining the first-order transition is a jump of the order parameter and internal energy at the critical temperature. In fig. \ref{V33-pw}, we see a double-peak structure of distributions for the energy and walls density. Such a structure is typical for a discontinues transition.

The valuable for us observation that the transition in the $\mathbb{Z}_2$ order parameter is crucial for the $SO(3)$ sector of the model. Instead of the $V_{3,2}$ crossover, the $V_{3,3}$ transition is a point where quantities in the $SO(3)$ sector change the thermal behavior from the $\sigma$-model behavior to the high-temperature one. In particular, this point corresponds to a jump of the vortices density and to a singularity of the topological susceptibility (fig. \ref{V33-vx}).

An inverse influence is also evident. In the absence of the $SO(3)$ sector, we would see the critical behavior of the Ising model universality class. So, if the order of the transition is the first, it is induced by fluctuations and topological defects of this sector.

\section{J$_1$-J$_3$ model}

\begin{figure}[t]
\includegraphics[scale=0.45]{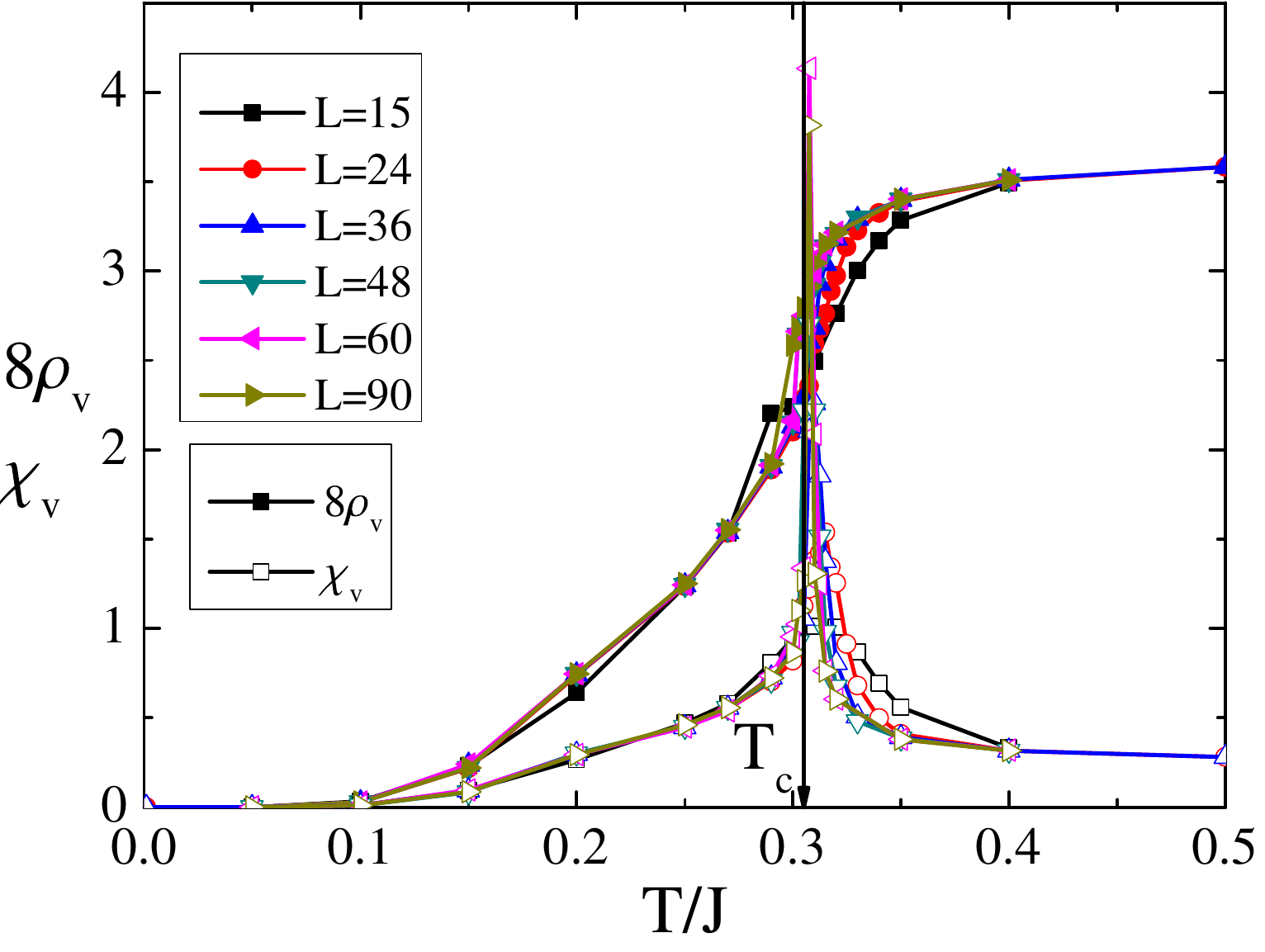}
\caption{\label{JJ-vx}Thermal properties of $\mathbb{Z}_2$-vortices in the J$_1$-J$_3$ model model.}
\end{figure}
\begin{figure}[t]
\includegraphics[scale=0.45]{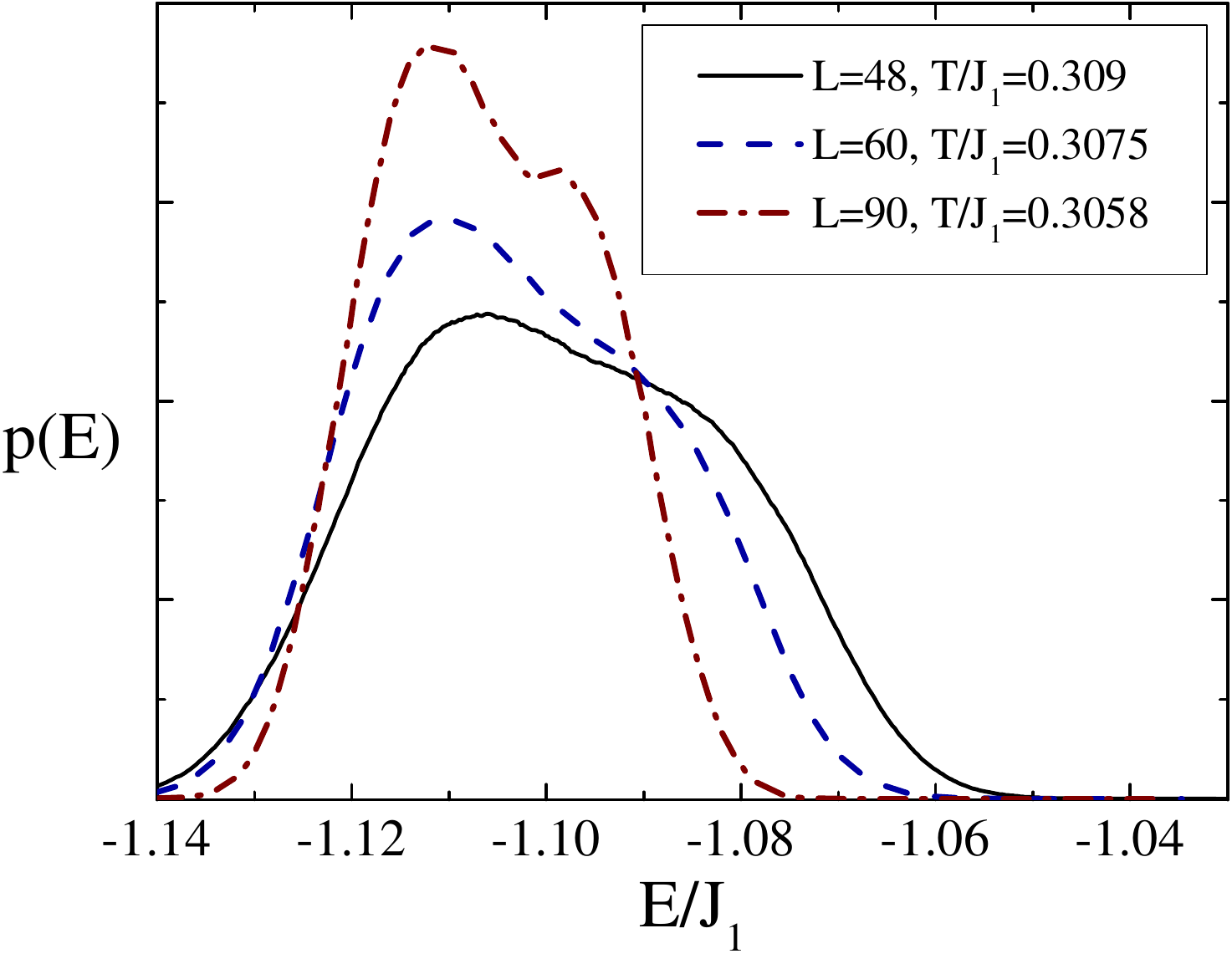}
\includegraphics[scale=0.45]{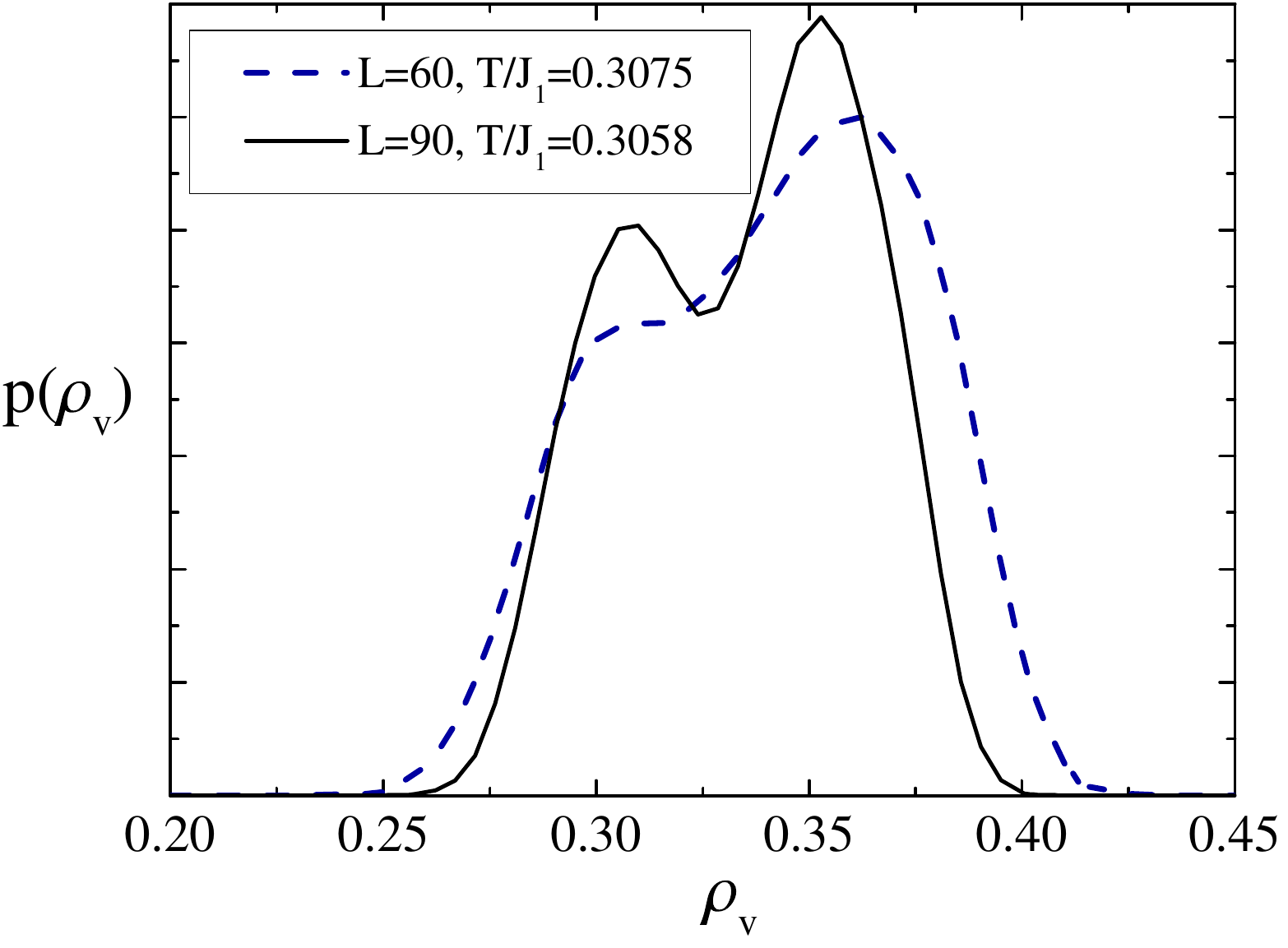}
\caption{\label{JJ-pv}Energy and vortices density distributions close to the transition point in the J$_1$-J$_3$ model model.}
\end{figure}

The J$_1$-J$_3$ model on a square lattice belongs to the same symmetry class as the $V_{3,3}$ Stiefel model. It is described by the Hamiltonian
\begin{equation}
    H=\sum_{\bfx,\mu}\left(J_1\bfS_{\bfx}\bfS_{\bfx+\bfe_\mu}+J_3\bfS_{\bfx}\bfS_{\bfx+2\bfe_\mu}\right),
    \label{JJmodel}
\end{equation}
with $J_1,\,J_3>0$. When the coupling constants $J_1$ and $J_3$ relate as $J_3<J_1/4$, the ground state is conventional N\'eel order with magnetic wave vector $\mathbf Q=(\pi,\pi)$. For $J_3>J_1/4$, the ground state has the planar incommensurate helical order with a wave vector $\mathbf Q=(q,q)$, where $\cos q=J_1/4J_3$.

This model has been intensively studied in the quantum case near the Lifshitz point $J_3=J_1/4$ in context of investigation a quantum spin-liquid state \cite{Sachdev04}. In the classical case, the model is also interesting. So, at non-zero temperatures and below the transition point, the model describes a phase with a chiral long-range order without a magnetic long-range or quasi-long-range orders. Such a phase is a classical spin liquid. (See \cite{Balents10,Staryh15} for a review.)

The model has been considered in three dimensions \cite{Sorokin11,Sorokin11-2,Sorokin14} as well as in two dimensions for the $N=2$ case \cite{Sorokin12}. In all these works, a single first order transition is found. However, in work \cite{Sachdev04}, a second order Ising transition is observed for the two-dimensional $N=3$ case, but our data favor the first order of a transition.

As a model of helimagnets, the J$_1$-J$_3$ model has a specificity affecting Monte Carlo simulations. At first, we deal with an incommensurate structure. Even if one chooses a helix pitch commensurate with a lattice size at the ground state, thermal effect increases a pitch, and a helix becomes incommensurate. Thus one have a troubles in choosing a periodic boundary conditions, and special algorithms should be used (see, e.g. \cite{Saslow92}). Secondly, a direct calculation of the helicity modulus becomes problematic (a reason have been discussed in \cite{Sorokin12,Sorokin12-2}). These difficulties are especially evident near the Lifshitz point, where a helix vector is large, but they can be partly ignored in a strongly frustrated case $J_3\geq J_1/2$. So, we consider the case $J_3=J_1/2$.

In contrast to the simple model of helimagnet \cite{Sorokin12-2}, The J$_1$-J$_3$ model has two chiral order parameters
\begin{equation}
    \bfk_{\bfx,\mu}=\frac{\bfS_\bfx\times\bfS_{\bfx+\bfe_\mu}}{\sin q_0},
\end{equation}
where $q_0$ is a helix vector length at zero temperature ($q_0=\frac{2\pi}{3}$ when $J_3=J_1/2$). One can return to notations of the $V_{3,3}$ model if chooses
\begin{equation}
    \bfk=\bfk_1, \quad \sigma=\mathrm{sign}(\bfk_1\cdot\bfk_2).
\end{equation}

We find the first order transition at temperature
\begin{equation}
    \frac{T_c}J=0.305(2).
\end{equation}
In order to determine an order of the transition, we use the same criteria as in the case of the $V_{3,3}$ model. In fig. \ref{V33-wx} the singularity of the topological (domain walls) susceptibility is shown, and it is clearly stronger than logarithmic. Fig. \ref{JJ-vx} demonstrates a singular behavior of $\mathbb{Z}-2$ vortices at the transition point. Jumps of the internal energy, order parameters and densities of topological defects are also observed. These results is partially shown in fig. \ref{JJ-pv}.

We also explore the possibility that the transition of a weak first-order (close to a second order), and a pseudo-scaling behavior may be observed. Our estimation of critical exponents indicate the distinct from the universality class of the Ising model. In particular, using the scaling relation $\alpha=2-2\beta-\gamma$, we obtain that the singularity of the specific heat and domain walls susceptibly corresponds to the exponent $\frac{\alpha}{\nu}\approx0.67$.
\begin{table}[h]
\caption{\label{table}Estimation of pseudo-exponents in the finite-size scaling (FSS). }
\begin{tabular}{c|c|c|c}
  \hline
  \hline
     &$\nu$&$\beta$&$\gamma$\\
  \hline
  Ising model & 1 & 0.125 & 1.75 \\
  This work & 0.72(5) & 0.07(1) & 1.38(10) \\
  I order FSS & 0.5 & 0 & 1 \\
  \hline
  \hline
\end{tabular}
\end{table}

\begin{figure}[t]
\includegraphics[scale=0.45]{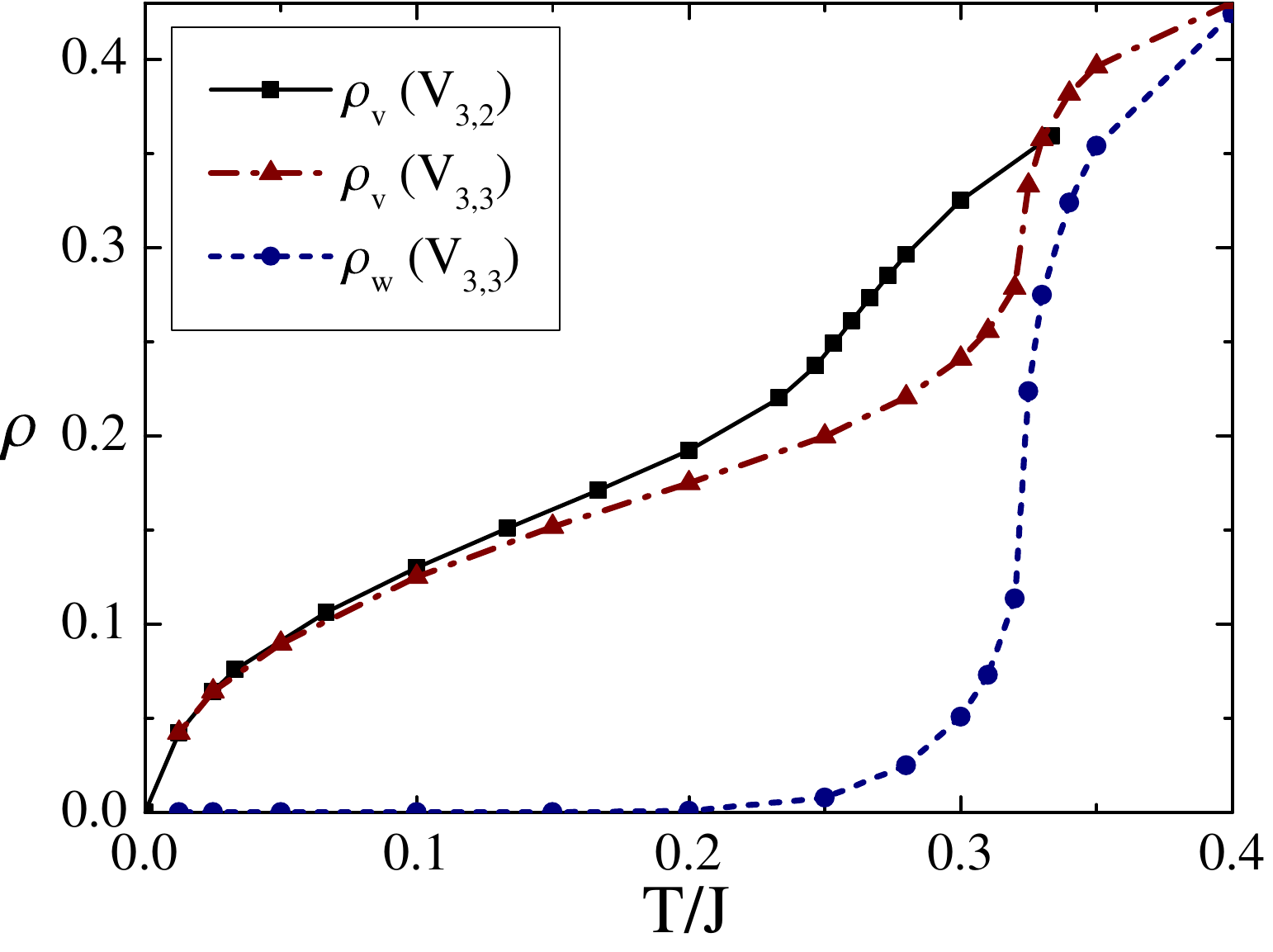}
\caption{\label{D-ro}Density of topological defects (walls and vortices) in the $V_{3,2}$ model with rescaled temperature and the $V_{3,3}$ model for $L=120$.}
\end{figure}

\section{Discussion}

We have considered the two spin systems, where the order parameter space is $\mathbb{Z}_2\otimes SO(3)$, and have found first order transitions. It is important, that we observe at the transition point a jump of density of topological defects (both domain walls and $\mathbb{Z}_2$-vortices). Such a jump absents at an Ising and BKT transition.

Comparing the model $V_{3,2}$ (without domain walls) with the $V_{3,3}$ model, we see the following picture (see fig. \ref{D-ro}). At low temperatures, when appearing of domain walls is suppressed, these models demonstrate the identical behavior of the vortices density. With temperature increasing, the crossover occurs in the $V_{3,2}$ model, and then the vortices density increases visibly. In the $V_{3,3}$ model, before the crossover occurs, domain walls start to appear in appreciable amounts. So then, a sharp increase of the walls and vortices densities is observed, and the first order transition occurs.

\begin{figure}[t]
\includegraphics[scale=0.35]{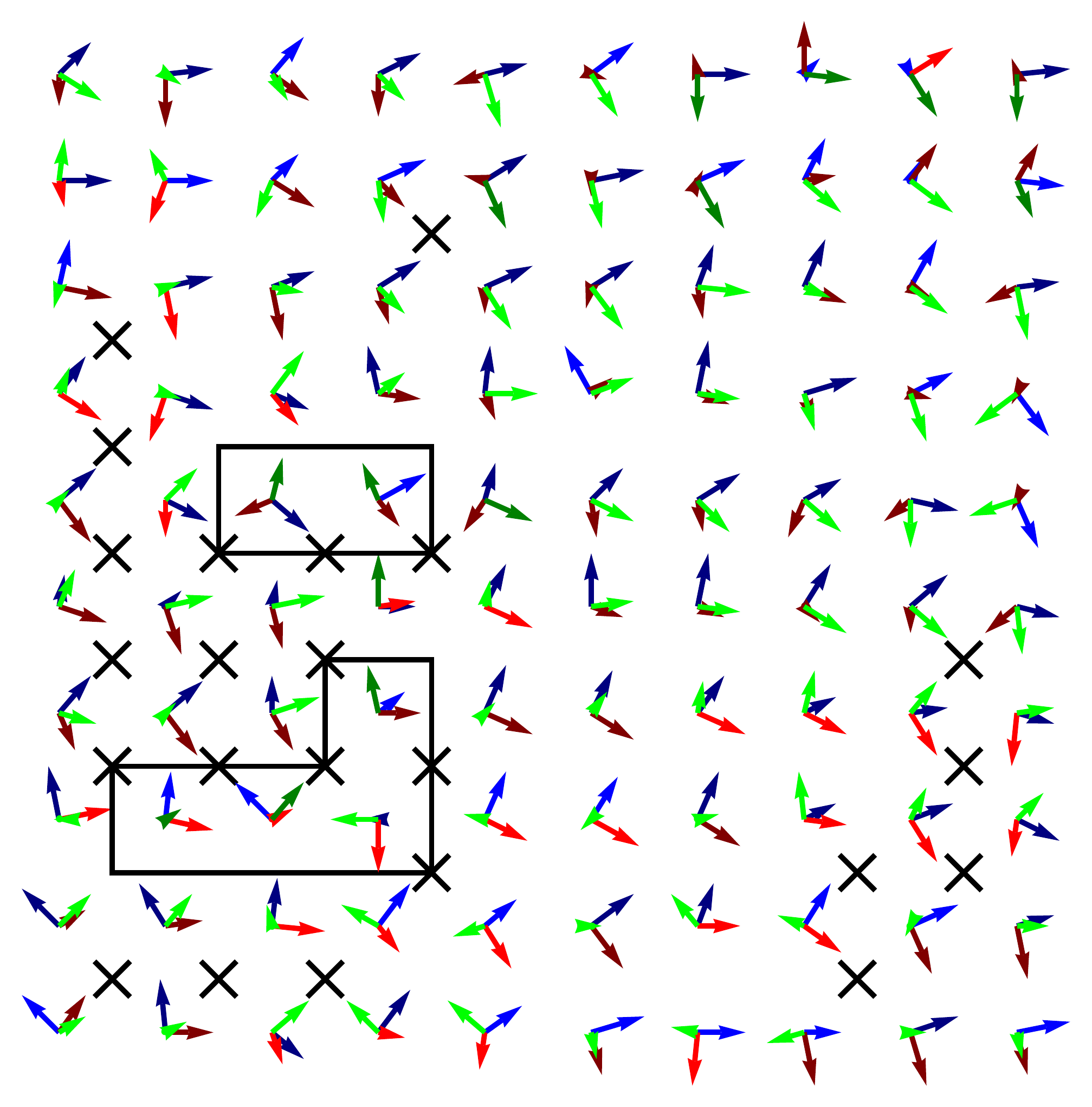}
\caption{\label{Vortex}Shot of a simulation the $V_{3,3}$ model at $T/J=0.3$.}
\end{figure}

The fact that the presence of $\mathbb{Z}_2$-vortices contributes to the domain walls density increasing, and vice versa the appearance of walls induces the vortices creation, we observe directly during the simulation process. Fig. \ref{Vortex} is the shot of a lattice fragment at $T/J=0.3$. It shows that domain walls and vortices are associated to each other.

The influence of $\mathbb{Z}_2$-vortices on the $\mathbb{Z}_2$ sector of the $V_{3,3}$ model leads to a change in the type of the critical behavior. In other words, the first order of the transition is induced by topological defects.

In the work \cite{Domenge08}, it has been discussed that the presence of $\mathbb{Z}_2$-vortices may not lead to a first order transition, if at the critical region vortices turn out heavier than walls (and their density is negligible), and the transition is driven by only domain walls, similar to the pure Ising model. We cannot exclude such a possibility. I.e. We do not exclude that a transition in a system with the $\mathbb{Z}_2\otimes SO(3)$ order parameter space is of a second order phase transition from the Ising model universality class.

\begin{acknowledgments}
This work is supported by the RFBR grant No 16-32-60143.
\end{acknowledgments}

\end{document}